\documentclass[aps,prl,twocolumn,showpacs,superscriptaddress,floatfix]{revtex4}
\usepackage{epsfig}\usepackage{amsmath}
\newcommand{\be}{\begin{equation}}
\newcommand{\bea}{\begin{eqnarray}}
\newcommand{\eea}{\end{eqnarray}}
\newcommand{\ee}{\end{equation}}

\begin{document}

\title{Robust creation of entanglement between ions in spatially separate cavities}

\author{Daniel E. Browne}\author{Martin B. Plenio}
\affiliation{QOLS, Blackett Laboratory, Imperial College London,
London SW7 2BW, UK} \author {Susana F. Huelga}
\affiliation{Division of Physics and Astronomy, University of
Hertfordshire, Hatfield, Herts AL10 9AB, UK}
%\maketitle
\date{\today}

\begin{abstract}
We present a protocol that allows the generation of a maximally
entangled state between individual atoms held in spatially
separate cavities. Assuming perfect detectors and neglecting
spontaneous emission from the atoms, the resulting idealized
scheme is deterministic. Under more realistic conditions, when the
the atom-cavity interaction departs from the strong coupling
regime, and considering imperfect detectors, we show that the
scheme is robust against experimental inefficiencies and yields
probabilistic entanglement of very high fidelity.

\end{abstract}
 \maketitle

The ability to reliably create entanglement between spatially
separate parties is of paramount importance for the actual
implementation of any quantum communication protocol \cite{Plenio
V 98} and is also a pre-requisite for distributed quantum
computation \cite{Cirac EHM 00}. Atoms or ions trapped inside
optical resonators provide a promising set up for demonstrating
the feasibility of quantum networking. Proposed ion trap quantum
gates \cite{Gates} allow the coherent processing of quantum
information stored in long lived electronic states. Indeed,
sequential gate operation allowed the first quantum algorithm to
be implemented in a linear ion trap \cite{Blatt 03}. Transferring
quantum information between distant sites could be achieved by
mapping the electronic degrees of freedom of the ions onto the
photonic degrees of freedom of the cavity, which can then be used
to transmit the quantum information to a distant site. First
experimental progress towards this direction has been recently
reported, demonstrating that individual ions can be positioned
inside an optical resonator achieving sub-wavelength position
control \cite{ionsincavities}. The next step would be the
controlled transfer of quantum information between electronic and
photonic qubits, which should then be mapped out of the cavity.

However, once the photon has left the resonator through one of its
highly reflecting mirrors it is not a straightforward task to feed
it into another cavity. Ingenious schemes using careful pulse
shaping have been devised to achieve this goal \cite{Repeater} but
their experimental implementation remains challenging. A
conceptually different approach consists of relaxing the condition
that the quantum information is transferred via a photon leaving
cavity A and entering cavity B.

\begin{figure}[htb]
\epsfysize=4.1cm
\begin{center}
\epsffile{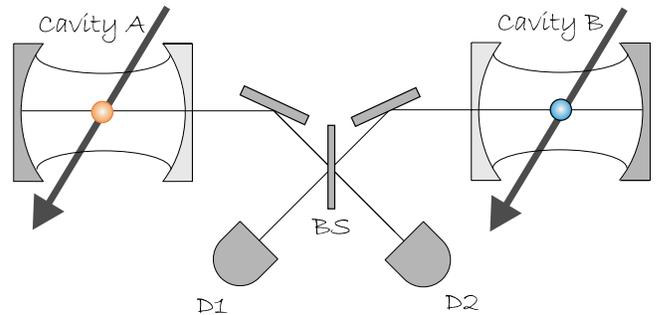}
\end{center}
\caption{\label{figurecavity} We consider a set-up in which
individual ions are trapped inside two spatially separated optical
cavities A and B.  Photons can leak out of the cavities and are
then mixed on a beam splitter BS and subsequently detected by
photodetectors $D_1$ and $D_2$.}
\end{figure}

Several schemes have been proposed for the generation entanglement between atoms, by detecting photons, in such a way that it is impossible to distinguish from which site they were emitted  \cite{Cabrillo,Bose KPV 99,Protsenko,feng,simon}.
For example, one could imagine a
setting as in Fig. \ref{figurecavity} where photons are allowed
to leave both cavities and are then mixed on a beam-splitter prior
to ordinary photon detection.
Which-path information is destroyed
in the beam-splitter and subsequent detection of a photon can then
lead to a projection of the electronic degrees of freedom of the
atom (which are entangled with the photonic degrees of freedom)
onto some maximally entangled state \cite{Bose KPV 99}.

In its original formulation, protocol \cite{Bose KPV 99} employed
sudden excitation of the ions, which, assuming otherwise perfect
experimental conditions, limited its efficiency to $50\%$.
Besides that constraint, two further problems would be difficult
to overcome in practice. The first and most serious one is that
the mapping between an ion and an optical cavity usually takes
place within the weak coupling regime, defined by the relationship
$g^2/\kappa\gamma\ll 1$. Here $g$ is the ion-cavity coupling for a
relevant set of atomic levels, $\kappa$ is the decay rate of the
optical cavity and $\gamma$ denotes the spontaneous decay rate on
the transition driven by the cavity mode. Within weak coupling, it
is very likely that the atom will suffer an incoherent spontaneous
emission, resulting in a photon leaving the cavity undetected to
the sides before the electronic degree of freedom has been mapped
onto the photonic degree of freedom. As it is very difficult to
detect a photon that is emitted to the sides of the cavities, this
event severely damages the quantum entanglement that one intends to
create.

Additionally, most protocols assume perfect detectors while in
practice they are generally not available. This problem is
compounded by the fact that in a number of setups for optical
cavities the mirrors possess considerable absorption which can be
as high as $50\%$ of the photons that are not reflected from the
cavity \cite{Hinds 03}. Therefore, any proposed scheme aimed to be
demonstrated with current technology needs to be highly
insensitive to detector inefficiencies. Many of the above
problems, such as weak coupling, poor detector efficiencies or
absorption in the mirrors, also occur, if one wishes to entangle
two ions in a single optical cavity by detecting photons as they
leak out of the mirrors. For this setting a number of schemes have
been put forward recently, see e.g. \cite{Plenio HBK 99,Sorensen M
02}. In \cite{Plenio HBK 99} an entangled state between the ions
is prepared conditional on the failure to detect a photon leaking
from the cavity. In practice the fidelity of the state decreases
very rapidly when one enters the weak coupling limit or when one
has imperfect detectors or absorption in the mirrors. The second
scheme \cite{Sorensen M 02} is more robust within the weak
coupling regime, but requires single photons pulses and suffers
strong loss of fidelity when faced with imperfect detectors or
absorption in the mirrors \cite{Duanaswell}.

In the following we present a scheme to entangle ions trapped
individually in spatially separated cavities which (i) succeeds
with $100\%$ probability under ideal conditions, (ii) allows the
achievement of high fidelity entanglement outside the strong
coupling regime upon the detection of a photon, (iii) is robust
against detector inefficiencies and absorption losses in the
cavity mirrors and (iv) can be adapted, with the same efficiency,
to entangle ions trapped in a single optical cavity.

\begin{figure}[htb]
\epsfysize=4.5cm
\begin{center}
\epsffile{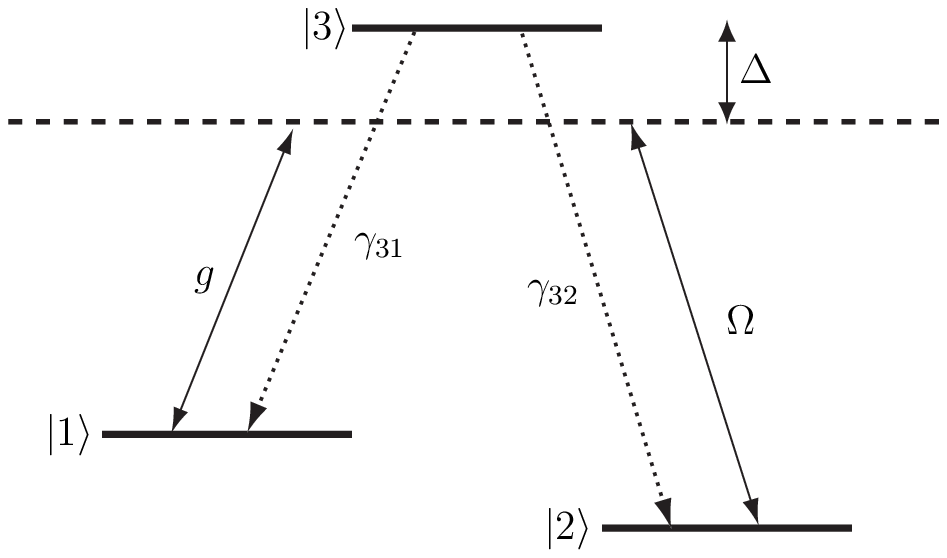}
\end{center}
\caption{\label{levelstructure} Internal level scheme of the ions.
A stable entangled state can be created when quantum information
is encoded in the lower two levels $|1\rangle$ and $|2\rangle$.
These two levels are coupled via the upper level $|3\rangle$
employing two fields that have the same large detuning $\Delta$ on
their respective transitions to the upper level $|3\rangle$. The
$|1\rangle\leftrightarrow |3\rangle$ transition couples to the
cavity mode while the $|2\rangle\leftrightarrow |3\rangle$
transition is driven by a strong
classical field. There may be spontaneous decay from $|3\rangle$
to levels $|1\rangle$ and $|2\rangle$ at rates $2\gamma_{31}$ and
$2\gamma_{32}$ respectively.}
\end{figure}

The method proposed here has its roots in the scheme presented in
\cite{Bose KPV 99}, where a teleportation protocol between two
cavities that employs the leakage of photons through the cavity
mirrors was discussed. The same method can also be used to
establish entanglement between the ions trapped in separate
cavities. We briefly describe this approach here to illustrate its
limitations and to motivate how to overcome them. Consider the set
up depicted in Fig. \ref{figurecavity}, where each cavity contains
a single trapped ion with an internal level structure which is
given in Fig. \ref{levelstructure}. Light that may leak through
the cavity mirrors is mixed on a $50/50$ beam splitter and
subsequently observed by photo-detectors. The qubit is represented
by the lower two energy levels which are coupled via a far detuned
Raman-like transition. In \cite{Bose KPV 99} it was envisaged that
the ions are both initially prepared in state $|2\rangle$. Then,
identical far-detuned classical light pulses are applied to both
ions such that, under ideal conditions, the state of the global
system is given by
\begin{eqnarray}
    |\psi_{tot}\rangle &=& \frac{1}{2}\biggl(|2_A,2_B\rangle|v_A,v_B\rangle
    -|1_A,1_B\rangle|p_A,p_B\rangle \nonumber\\
    &+&
    i(|2_A,1_B\rangle|v_A,p_B\rangle+|1_A,2_B\rangle|p_A,v_B\rangle)\biggr),
\end{eqnarray}
where $|v_A\rangle$ represents the vacuum state in cavity A and
$|p_B\rangle$ denotes the one-photon Fock state in cavity B.
Following this pulse, one waits to allow photons to leak through
the cavity mirrors, mix at the beam splitter and reach the
detectors.
 If a single click occurs, then the system is
projected onto one of the two entangled states $\{
(|2_A,1_B\rangle+|1_A,2_B\rangle)/\sqrt{2},(|2_A,1_B\rangle-|1_A,2_B\rangle)/\sqrt{2}\}$.
If no photon is detected or two photons are detected, then the
ions are projected onto a product state and the procedure has
failed.

Apart from the sensitivity of this scheme to losses due to
spontaneous emission and detector inefficiencies, the procedure
fails even under ideal conditions in $50\%$ of the cases. The
reason for the $50\%$ failure rate of the scheme is that we excite
both ions suddenly, which leads to a very high probability for the
two photon detection event, leaving the ions in a product state.
Furthermore, the scheme is not robust to spontaneous decay of the
ions (this is particularly relevant in the weak coupling regime)
or detector inefficiencies. If spontaneous emission occurs in one
of the ions, the photon emitted will escape undetected, the
detection of one photon in the photo-detectors after this event
will then lead to the generation of the product state
$|1_A,1_B\rangle$, not the desired entangled state. Secondly, if
the detector is inefficient, then only one of the photons of a two
photon event, i.e. a failure of the protocol, might be detected.
Both these cases will therefore lead to a potentially severe
reduction in fidelity of the states produced by the protocol.

However, an additional ingredient can make this scheme highly
robust against any of these error sources. This is achieved by
relaxing the condition of sudden excitation of the two ions and
replacing it by more gentle driving. In the following we will show
that weak driving, under ideal conditions, allows the scheme to
succeed with arbitrarily high probability (and unit fidelity).
The Hamiltonian of the combined ion-cavity system, with ion
internal level structure as given in Fig. \ref{levelstructure},
in a suitable interaction picture and setting $\hbar=1$, a convention we will use throughout this paper, is given by
\begin{eqnarray}
    H &=& \sum_{i=A,B}\hspace{-1mm} \biggl(\Delta |3\rangle_i{}_i\langle 3| +
    g |3\rangle_i{}_i\langle 1|c_i + g |1\rangle_i{}_i\langle
    3|c_i^{\dagger} \nonumber\\
    && +
    \Omega |3\rangle_i{}_i\langle 2| + \Omega |2\rangle_i{}_i\langle
    3|\biggr),
\end{eqnarray}
where we have assumed that the two ions are subjected to identical
laser fields on the  $|2\rangle\leftrightarrow
|3\rangle$ transition and that this laser field has the same
detuning as the cavity field that couples to the
$|1\rangle\leftrightarrow |3\rangle$ transition. The annihilation
and creation operators for the cavity photons are denoted by $c_i$
and $c_i^{\dagger}$. The upper level $|3\rangle$ of both ions can
decay to levels $|1\rangle$ and $|2\rangle$ with a rate of $2\gamma_{31}$ and  $2\gamma_{32}$, respectively.
 Each of the
cavities has a decay rate $2\kappa$. The full master equation for
the density operator $\rho$ then takes the form
\begin{eqnarray}
    \dot\rho &=& -i(H_{\text{eff}}\rho -
    \rho H_{\text{eff}}^{\dagger}) + 2\kappa \sum_{i=A,B} c_i\rho
    c_i^{\dagger}\nonumber \\
    && + \sum_{i=A,B} 2\gamma_{31} |1\rangle_i{}_i\langle 3|\rho |3\rangle_i{}_i\langle
    1|\nonumber
\\
    && + \sum_{i=A,B} 2\gamma_{32} |2\rangle_i{}_i\langle 3|\rho |3\rangle_i{}_i\langle
    2|,
\end{eqnarray}
where we have defined the effective non-Hermitian Hamiltonian
\begin{eqnarray}
    H_{\text{eff}} = H - i\kappa \sum_{i=A,B} c_i^{\dagger}c_i -i \left(\gamma_{31}+\gamma_{32}\right)\hspace{-1mm} \sum_{i=A,B}\hspace{-1mm}|3\rangle_i{}_i\langle 3| .
\end{eqnarray}
This effective Hamiltonian will be used in the quantum jump
approach to describe the system dynamics under the condition that
neither a spontaneous emission nor a cavity photon have been
detected \cite{Plenio K 98}.

For the sake of simplicity we now consider the system in the
strong coupling limit, by setting $\gamma_{31}=\gamma_{32}=0$. We will later relax this
condition to show that our method also works in the weak coupling
limit. The requirement of weak driving means that the condition
$\frac{g\Omega}{\Delta} \ll \kappa$ is satisfied. Intuitively
this implies that the rate of transitions between levels $|1\rangle$
and $|2\rangle$ of the ions will be weaker than the cavity decay. This in
turn implies that the population in level $|1\rangle$ of the atoms
will be small, unless a photon is detected. Indeed, after adiabatic
elimination of the upper level $|3\rangle$ we obtain that the weak driving
dynamics is governed by the new master equation  \cite{pellizzari},
\begin{eqnarray}
    \dot\rho &=& -i(H_{\text{ad}}\rho -
    \rho H_{\text{ad}}^{\dagger}) + 2\kappa \sum_{i=A,B} c_i\rho
    c_i^{\dagger},
\end{eqnarray}
where we have defined
\begin{equation}\label{eq:had}
\begin{split}
    H_{\text{ad}} &= \sum_{i=A,B} \biggl[\frac{g\Omega}{\Delta}
    \left(|2\rangle_i{}_i\langle 1|c_i + h.c.
    %|1\rangle_i{}_i\langle 2|c_i^{\dagger}
    \right)
+\frac{g^2}{\Delta}|1\rangle_i{}_i\langle 1|\\&\qquad\mbox{}
+\frac{\Omega^2}{\Delta}|2\rangle_i{}_i\langle 2|
- i\kappa c_i^{\dagger}c_i\biggr] \; .
\end{split}
\end{equation}
Under the condition that no detection has been registered, the
time evolution is governed by $H_{ad}$. Given an initial state
$|2_A,2_B\rangle|v_A,v_B\rangle$, the state of the systems will
quickly approach the form
\begin{eqnarray}
    |\psi\rangle &=& |2_A,2_B\rangle|v_A,v_B\rangle +
    x\biggl(|2_A,1_B\rangle|v_A,p_B\rangle\\&&\mbox{}+|1_A,2_B\rangle|p_A,v_B\rangle\biggr)+O(x^2),
    \nonumber
\end{eqnarray}
where $x\cong-i\frac{g\Omega}{\Delta\kappa}$. Therefore, the rate $R$
 at which one observes photons in one of the detectors is
proportional to
$R\cong4\kappa\left(\frac{g\Omega}{\Delta\kappa}\right)^2$. Note that  in this regime this rate is essentially unaffected by the level shifts in $H_{\text{ad}}$, thus it is not necessary for these to be compensated for. If
one of the photo-detectors clicks, then a maximally entangled
state has been prepared to a high precision and one switches off
the lasers so that the entangled state is then preserved as the
ions decouple from the cavity. The mean time before the first
detection event will be
\begin{equation}\label{tav}
T_{\text{av}}\cong\frac{\Delta^2\kappa}{4(g\Omega)^2}.\end{equation}
 In such
a time interval, there is a small probability that two photons are
detected, however this probability scales as
$|x|^4 \ 2\kappa
T_{\text{av}}
    \cong \frac{1}{2}\left(\frac{g\Omega}{\Delta\kappa}\right)^2
$ and can therefore be made arbitrarily small in the limit of
large detuning. We then observe that we can prepare a perfectly
entangled state with arbitrarily high fidelity if we choose a
sufficiently high detuning or sufficiently weak coupling strengths
$g$ and $\Omega$. As a consequence, by choosing a detuning that is
very large, i.e. driving the $|1\rangle\leftrightarrow |2\rangle$
transition very slowly, we can ensure that any detection event is
linked to a {\em single} photon and that therefore the fidelity of
the prepared state will be very close to unity. This demonstrates
our first claim that the scheme can achieve perfect fidelity and
unit success probability.

However, this result is still only valid in the strong coupling
limit as we have so far neglected the effect of finite
$\gamma_{31}$ and $\gamma_{32}$. The strong coupling limit is not
easy to achieve experimentally and it would be desirable to have a
procedure that generates a very high fidelity entangled state even
away from this limit, ideally with reasonable success probability.
In the following we will demonstrate that our scheme can also
successfully produce high fidelity maximally entangled states
outside the strong coupling limit  where we allow
$\frac{g^2}{\kappa\gamma_{31}} \approx 1$ or even
$\frac{g^2}{\kappa\gamma_{31}} \ll 1$ and $\gamma_{32}$ is allowed
to be non-vanishing.
 In fact, in this
case we can still achieve very high fidelities at good success
rates. This observation is confirmed by a numerical simulation for
the following choice of parameters $\Omega=g, \kappa=10g,
\gamma_{31}=\gamma_{32}=0.1g$ and
$\Delta=20g$ and a waiting time of $T=100/g$ we obtain numerically
a fidelity of $F=0.98$ with a success probability of the scheme of
$p\cong0.1$. Indeed, analytically the success probability is
approximately given by
\begin{equation}
    p_{suc} \cong \left(\frac{g\Omega}{\Delta\kappa}
    \right)^2 4 \kappa T
\end{equation}
where $T$ is the time one is willing to wait for the first
detection. If no click is observed after this time, the experiment
is deemed a failure and the systems will be e-prepared with both
atoms in state $|2\rangle$.

The previous considerations still assume that the detection
efficiency for photons that leak out of the cavity is unity.
However, there are important sources of losses in experiments that
make this assumption unrealistic. Firstly, there may be absorption
in the mirrors themselves \cite{Hinds 03} and secondly the
detectors may only have a finite efficiency. A scheme that can
work in a practical environment should therefore also be robust
against detector inefficiencies. Fortunately, the present method
exhibits exactly such a robustness. In terms of the detector
efficiency $\eta$ we find that the success probability simply
scales linearly as
\begin{equation}
    p_{\text{suc}} \cong \left(\frac{g\Omega}{\Delta\kappa}
    \right)^2 4\kappa \eta T .
\end{equation}
With falling detector efficiency, the fidelity of the resulting
state will decrease, because it will now contain an admixture from
events where two photons have been emitted from the cavities, but
only a single one has been detected. Indeed, from this argument
one expects a weak linear reduction of the fidelity.

\begin{figure}[htb]
\epsfxsize=8.7cm
\begin{center}
\epsffile{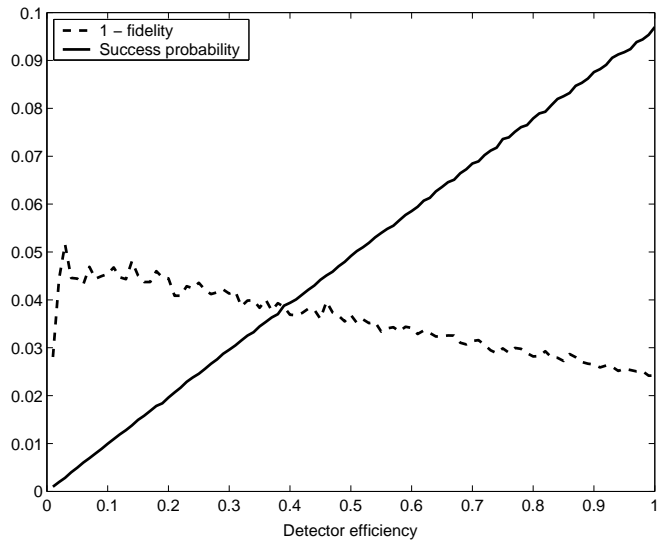}
\end{center}
\caption{\label{effprob} The success probability (solid line) and
the deviation from unit fidelity, are plotted versus the detector
efficiency for $\Omega=g, \kappa=10g, \gamma_{31}=\gamma_{32}=0.1g$ and $\Delta=20g$ and a waiting
time of $T=100/g$. The plot has been obtained from a quantum jump
simulation of the exact dynamics given by Eq. 4. Each point is the
result of an average over $10^6$ runs of the scheme.}
\end{figure}

In Fig. \ref{effprob} we have plotted both the success probability
and the achieved error (1-fidelity) for fixed $\Omega=g,
\kappa=10g, \gamma_{31}=\gamma_{32}=0.1g$ and
$\Delta=20g$ and a waiting time of $T=100/g$ against the detector
efficiency. This figure confirms the approximate analytical
formulas presented above and underlines that our scheme is robust
against variations in the detector efficiencies.
%%%%%Dark Counts

A further experimental imperfection which must be considered is
the presence of ``dark counts'', i.e. when the detector fires
although no light is incident upon it. Clearly, this will degrade to
some extent the performance of all schemes which rely on a single
detector click to generate an entangled state, and will lead, in
general, to a loss in fidelity of the state produced. However, in
the present scheme, the time-window in which a click due to a
photon should occur is far shorter than the mean time between dark
counts. For example, in \cite{Buttler} a dark count rate of
approximately 1400 s$^{-1}$ is reported, thus the mean time
between dark counts is on the order of ms. In the optical regime,
the atom-cavity coupling $g$, detuning $\Delta$, cavity decay rate
$\kappa$ and the coupling with the classical field $\Omega$ will
all be at least on the order of MHz \cite{Kimble}. Thus, using Eq.
\eqref{tav} one can estimate that $T_\text{av}$, the mean time
before a proper click occurs in this scheme, is on the order of
$\mu$s. Since the time-window for detection in this scheme can
thus be made much smaller than the mean time between dark counts,
their effect  on this scheme can be made very small.
%%%%%% Dark counts

So far we have considered the case where we were faced with the
task of entangling two spatially separated ion-cavity systems. The
key ingredient in the detection scheme was the beam-splitter that
erased the which-path information from the system, so that a
photo-detection event could lead to entanglement between the
cavities. However, the above method could also be used to entangle
two ions trapped in a single cavity whose decay is monitored by a
single photo-detector if the system is set-up such that the
detection of a photon does not provide any information about which
ion the photon was emitted from.

In summary, we have presented here an approach that, under ideal
conditions, allows for the deterministic generation of perfect
entanglement between individual ions in distant cavities. In the ideal scenario, the unit success probability also allows for the
generalization of this scheme to the direct implementation of
quantum gates. The
scheme can be adapted to entangling multiple ions in a single
optical cavity. Most importantly, the scheme is  robust to realistic experimental
imperfections, and in particular it allows for the probabilistic
generation of high fidelity entanglement when operated within the
weak coupling limit and monitored by inefficient detectors.

\acknowledgements The authors would like to thank P. L. Knight for helpful advice. This work was supported by the US-Army through
grant DAAD 19-02-0161, the Nuffield Foundation and Hewlett Packard
Ltd. via an EPSRC CASE award.

%\end{multicols}


\begin{thebibliography}{99}
\bibitem{Duanaswell} After completion of  this manuscript, a further very
promising scheme was suggested by Duan and Kimble \cite{Duan
K 03}, which provides high fidelity entanglement in the weak
coupling limit and in the presence of inefficient detectors. It
does however have some drawbacks as the inherent maximal success
rate is $50\%$ and this rate decreases quadratically with detector
efficiency. While this may not be a very serious impediment for
the single-shot creation of entanglement, it is problematic when
one attempts to extend the scheme towards the implementation of
quantum gates and entanglement purification protocols.
\end{thebibliography}
\end{document}